\documentclass[manuscript,screen, sigconf]{acmart}
\usepackage{subcaption}
\usepackage{colortbl}
\usepackage{balance}
\AtBeginDocument{%
  \providecommand\BibTeX{{%
    \normalfont B\kern-0.5em{\scshape i\kern-0.25em b}\kern-0.8em\TeX}}}

\copyrightyear{2024} 
\acmYear{2024} 
\setcopyright{rightsretained} 
\acmConference[CHI EA '24]{Extended Abstracts of the CHI Conference on
Human Factors in Computing Systems}{May 11--16, 2024}{Honolulu, HI, USA}
\acmBooktitle{Extended Abstracts of the CHI Conference on Human Factors in
Computing Systems (CHI EA '24), May 11--16, 2024, Honolulu, HI, USA}
\acmDOI{10.1145/3613905.3650882}
\acmISBN{979-8-4007-0331-7/24/05}

\begin{document}

\title[Justice and Fairness Content Moderation on Social Media]{Content Moderation Justice and Fairness on Social Media: Comparisons Across Different Contexts and Platforms}

\author{Jie Cai}
\affiliation{%
  \institution{Pennsylvania State University}
  \city{University Park}
  \country{USA}}
\email{jie.cai@psu.edu}
\orcid{0000-0002-0582-555X}

\author{Aashka Patel}
\affiliation{%
  \institution{New Jersey Institute of Technology}
  \city{Newark}
  \country{USA}}
\email{asp263@njit.edu}
\orcid{0000-0003-4566-2537}

\author{Azadeh Naderi}
\affiliation{%
  \institution{New Jersey Institute of Technology}
  \city{Newark}
  \country{USA}}
\email{an57@njit.edu}
\orcid{0000-0002-0732-9912}

\author{Donghee Yvette Wohn}
\affiliation{%
  \institution{New Jersey Institute of Technology}
  \city{Newark}
  \country{USA}}
\email{yvettewohn@gmail.com}
\orcid{0000-0001-5583-4430}

\renewcommand{\shortauthors}{Cai et al.}

\begin{abstract}
Social media users may perceive moderation decisions by the platform differently, which can lead to frustration and dropout. This study investigates users’ perceived justice and fairness of online moderation decisions when they are exposed to various illegal versus legal scenarios, retributive versus restorative moderation strategies, and user-moderated versus commercially moderated platforms. We conduct an online experiment on 200 American social media users of Reddit and Twitter. Results show that retributive moderation delivers higher justice and fairness for commercially moderated than for user-moderated platforms in illegal violations; restorative moderation delivers higher fairness for legal violations than illegal ones. We discuss the opportunities for platform policymaking to improve moderation system design.

\end{abstract}

\ccsdesc[500]{Human-centered computing~Empirical studies in collaborative and social computing}
\ccsdesc[500]{Human-centered computing~Empirical studies in HCI}

\keywords{Social Media, Content Moderation, Justice and Fairness, Platform Governance, Policymaking}

\maketitle
\section{Introduction}
To fight harmful content and maintain a safe online space, platforms use algorithms to automatically or human labor to remove harmful content and sanction offenders manually \cite{gillespie_custodians_2018}, which is termed ``content moderation.'' Content moderation is defined as ``the practice of screening of user-generated content (UGC) posted to Internet sites, social media platforms and other online outlets, to determine the appropriateness of the content for a given site, locality, or jurisdiction'' \cite{roberts_content_2017}. Social media platforms have developed complex and platform-specific content moderation policies to regulate harmful content \cite{pater_characterizations_2016}. This paper uses the term ``policies'' to indicate formalized statements such as rules, standards, terms of services, and community guidelines. These content moderation policies guide human moderators and algorithms to remove offensive content and sanction offenders \cite{cai_moderation_2021}. Many users complain their content has been removed by the platform without a clear explanation and frequently express confusion about what action has triggered a moderation action or account suspension \cite{lyu_i_2024, ma_im_2022, elmimouni_shielding_2024}, leading users to feel frustrated and leave \cite{myers_west_censored_2018}. 

As moderation strategies vary across different contexts and platforms, users may perceive different levels of justice and fairness in the moderation decision. Many HCI scholars have explored the platform's moderation mechanisms from various perspectives, such as human labor \cite{steiger_psychological_2021} and transparency \cite{ma_how_2023, ma_defaulting_2023}. From the end-user perspective, some work explores the bystander effect \cite{blackwell_when_2018}, and others explore victims of harmful content \cite{schoenebeck_youth_2021,xiao_sensemaking_2022}. For example,  recent work shows that explanation of content removal increases a user's perceived fairness and engagement in the community again, and a user’s perceived education and interaction are more effective than blocking \cite{cai_what_2019}, suggesting that removal explanation is positively associated with users’ perceived justice and fairness.

In line with working to understand the user's perception of moderation decisions, we ran an online experiment to understand how the explanation in different contexts with different harmful content can influence users’ perception of fairness and justice. We extend the line of research about justice and fairness in content moderation. The findings can potentially benefit social media platforms by furthering their understanding of how their users perceive moderation decisions and enabling them to better current moderation practices with policymaking. 

\section{Related Work}

\subsection{Dependent Variables: Justice and Fairness}
Computing systems are embedded with different biases regarding context, user populations, and technical constraints, further leading to unfairness and injustice in use \cite{friedman_bias_1996, smith_many_2023}. Content moderation as a sociotechnical process can privilege normalized groups while pushing other user groups with different races, genders, and religions, to the margins \cite{das_jol_2021}. 

The justice lens from criminal justice systems has been applied to the content moderation domain; if users perceive moderation decisions to lack justice and fairness, they are likely to stop abiding by the rules of the platforms or even seek their own ways to punish the offenders, resulting in punishment that may be indeterminate, uncalibrated, or inaccurate \cite{blackwell_when_2018}. Moderation decisions considering different contexts and offender’s characteristics can potentially increase the perceived fairness and justice \cite{cai_after_2021,schoenebeck_drawing_2021}. As such, in order to keep online communities safe and civil, we aim to assess users’ perceptions of justice and fairness on social media platforms. 

Justice is the perceived adherence to rules that reflect appropriateness in decision contexts \cite{colquitt_justice_2015}, including consistency, accuracy, bias suppression, and correctability \cite{thibaut_procedural_1976}. Justice has been discussed broadly in online harassment and platform governance \cite{cai_after_2021,kou_punishment_2021,schoenebeck_reimagining_2021}. In our study, we center the general users'/bystanders' perspective of justice in content moderation. We view it as adherence to rules of conduct regarding online content violation and moderation strategies applied. Fairness is an individual's moral evaluation of the rules of conduct \cite{colquitt_measuring_2015}. Fairness is an important factor for online content moderation. For instance, providing explanations for content removal can increase an offender's perceived fairness and desire to keep participating in the community \cite{jhaver_did_2019}. In this study, users’ perceived fairness of moderation decisions is the users’ moral evaluations of rule enforcement regarding online content violations and moderation strategies.

\subsection{Independent Variables: Platform Types, Violation Types, Moderation Types}

\subsubsection{Commercially moderated Versus User-moderated Platforms}
While specific platforms practice a bottom-up governance model that relies on community members to enforce policies, others practice top-down governance in which “officials implement a relatively detailed set of rules over a given community” \cite{bradford_ben_report_2019}. Examples of such platforms include Facebook and Twitter, where online harassment is dealt with centrally rather than relying on volunteer moderators \cite{cai_moderation_2021}. Commercial content moderators for these platforms are paid and contingent \cite{roberts_content_2017}. Users of such platforms believe that the company should bear more responsibility for content moderation rather than having the responsibility fall on the users themselves \cite{patel_user_2021}. On the other hand, with user-moderated platforms such as Twitch, creators of the user-generated content have moderating privileges and can appoint their followers to be additional moderators \cite{wohn_volunteer_2019}. Often, volunteer moderators find personal meaning in their roles \cite{seering_who_2022} and want to strengthen their online communities by guiding and developing offenders rather than simply “cleaning up” misbehavior \cite{seering_moderator_2019, wohn_volunteer_2019}. Overall, there are apparent differences between how commercial and user moderators treat offenders \cite{cook_commercial_2021}; consequently, the different treatment may affect users’ perceived justice and fairness of the punishment. As such, we ask our research question:

\begin{itemize}
\item 1) Will users’ perceived justice and fairness of online moderation decisions be higher or lower for commercially moderated versus user-moderated social media platforms?
\end{itemize}

\subsubsection{Illegal versus Legal Content Scenarios}

Social media platforms face increasing pressure from users and lawmakers to “clean up their platforms'' \cite{leetaru_is_2018}. The legality of content posted online matters for some users, and these users generally believe that platforms should not have the power to remove this content “as long as their posts are not illegal and do not incite illegal assembly, destruction of property or violence”  \cite{morici_facebook_2021}. This study considers illegal violations as content-inducing crimes or public safety concerns. Users believe social media platforms should have the ability and fairness to remove and moderate illegal material, such as child pornography, from their platforms \cite{langvardt_regulating_2018, jiang_understanding_2021}.

Online abuse such as racial slurs, bullying, sexual harassment, spam, trolls, and hate speech toward a specific group or individuals are also considered violations \cite{blackwell_classification_2017, jiang_understanding_2021}. This abuse is mainly handled by the platform or even the specific entities like end-users and human moderators \cite{cai_moderation_2021}. In this study, by community guidelines of various social media platforms, we consider this online abuse targeting a specific group or individual without severe public impact as a legal violation. These violations are broadly defined and contingent on different platforms and communities. Because the belief that illegal content should be moderated on social media platforms echoed in existing literature and media articles, we developed the following hypotheses:
\begin{itemize}
\item H2a. Perceived justice is higher in illegal compared to legal content scenarios.
\item H2b. Perceived fairness is higher in illegal compared to legal content scenarios.
\end{itemize}

\subsubsection{Restorative Versus Retributive Moderation Strategies}
Retributive justice is “a theory of punishment in which individuals who knowingly commit an act deemed morally wrong receive a proportional punishment for their misdeeds” \cite{blackwell_when_2018}. Restorative justice is “a process whereby all the parties with a stake in a particular offense come together to collectively resolve how to deal with the aftermath of the offense and its implications for their future” \cite{messmer_restorative_1992}. 
Generally, platforms remove offensive content and the offender from their communities. This content moderation practice is believed to echo the American criminal justice system and retributive justice \cite{hasinoff_promise_2020}. Through restorative justice, the offenders are meant to acknowledge their wrongdoings, accept responsibility for their transgressions, and demonstrate remorse \cite{schoenebeck_drawing_2021}. As such, through restorative content moderation, social media platforms can build healthier, resilient, and long-term online communities \cite{hasinoff_promise_2020}.


We adapt the retributive and restorative perspectives and divide the moderation strategies into retributive and restorative moderation, similar to the moderation styles punishing and nurturing \cite{jiang_trade-off-centered_2023}. Retributive moderation strategies in this study refer to banning offenders for rule-breaking and incapacitating offenders’ community participation. Restorative moderation strategies in this study refer to light punishment compared with the same violation (e.g., warning offenders with rule explanation and potential consequences if offenders keep behaving similarly) to maintain the community. Users of social media platforms may prefer one form of moderation strategy over the other. As such, we developed the following hypotheses:

\begin{itemize}
\item H3a. Perceived justice is higher for retributive compared to restorative moderation strategies.
\item H3b. Perceived fairness is higher for retributive compared to restorative moderation strategies.
\end{itemize}

\begin{table*}[htbp]
\small\sf\centering
\caption{Scenario Descriptions in the Experiment}
\label{tab:scenario}
\resizebox{\textwidth}{!}{%
\begin{tabular}{p{4cm}p{4cm}p{4cm}p{4cm}}
\toprule
Scenario 1a   & Scenario 1b  & Scenario 1c & Scenario 1d  \\
\midrule
a restorative moderation strategy for illegal content on a  user-moderated platform & a restorative moderation strategy for   illegal content on a commercially moderated platform & a retributive moderation strategy for illegal content on a   user-moderated platform & a retributive moderation strategy for illegal content on a   commercially moderated platform  \\
\midrule
Scenario 2a  & Scenario 2b  & Scenario 2c & Scenario 2d \\
\midrule
a restorative moderation strategy for   legal content on a user-moderated platform   & a restorative moderation strategy for legal content on a   commercially moderated platform   & a retributive moderation strategy for   legal content on a user-moderated platform   & a retributive moderation strategy for   legal content on a commercially moderated platform  \\
\bottomrule
\end{tabular}
}
\end{table*}

\section{Methods and Materials}

We choose Twitter (now ``X'') and Reddit as our research context. Twitter is a typical platform applying commercial moderation, while Reddit is known for user moderation. Both platforms provide thread-style communication with sharing and commenting features. The similar affordances and the different moderation policies indicate they fit well for design experiments with various scenarios. 

\subsection{Online Experiment Deployment}
We conducted an online experiment with participants recruited from Qualtrics. We limited the recruitment of participants to the United States to ensure that all participants had a similar understanding of the topic. We also set the gender as 50\% male and 50\% female to control the gender bias.  Our sample size was 200 (100 with either legal or illegal scenarios). To ensure our sample was as representative as possible, we asked Qualtrics to set these quota guidelines. If potential participants were under 18 and did not utilize Twitter and/or Reddit, they were terminated from the online experiment. If online experiment takers answered either “I will not provide my best answers” or “I can’t promise either way” to our question “Do you commit to providing your thoughtful and honest answers to the questions in this online experiment?”, they were also terminated. In the consent form at the beginning of the survey, we stated that participants would be exposed to violent and controversial content with relevant mental health resources to mitigate the negative impact of this study.  The median time for participants to complete the online experiment was 7.4 minutes. If participants completed the online experiment, they were compensated per their agreement with Qualtrics, the panel provider. We paid Qualtrics \$5.00 (USD) for each participant. Our participants were 18 to 65 or older, mostly between 30-49 (50.5\%)  and White (69\%). 

\begin{table*}[t]
\small\sf\centering
\caption{Descriptive Statistics for Each Group and Scenario in Terms of Perceived Justice}
\label{tab:justice}
\begin{tabular}{llllll}
\toprule
                & Justice-a             & Justice-b             & Justice-c             & Justice-d             & Justice-overall       \\ \midrule
                & Scenario\_1a          & Scenario\_1b          & Scenario\_1c          & Scenario\_1d          &                       \\ 
Group1: Illegal & \textit{M} = 3.32, \textit{SD}   = 1.30 & \textit{M} = 3.19, \textit{SD}   = 1.29 & \textit{M} = 3.93, \textit{SD}   = 1.35 & \textit{M} = 4.48, \textit{SD}   = 1.10 & \textit{M} = 3.73, \textit{SE}   = 0.11 \\ \midrule
                & Scenario\_2a          & Scenario\_2b          & Scenario\_2c          & Scenario\_2d          &                       \\ 
Group2: Legal   & \textit{M} = 3.60, \textit{SD}   = 1.33 & \textit{M} = 3.77, \textit{SD}   = 1.34 & \textit{M} = 3.45, \textit{SD}   = 1.49 & \textit{M} = 3.66, \textit{SD}   = 1.44 & \textit{M} = 3.62, \textit{SE}   = 0.11 \\ \midrule
                & \textit{p} = .136             & \textit{p} = .002             & \textit{p} = .019             & \textit{p} <  .001     & \textit{p} = .487             \\ \bottomrule
\end{tabular}%
\end{table*}

\begin{figure*}[ht]
\centering
\begin{subfigure}{0.33\textwidth}

  \includegraphics[width= \linewidth]{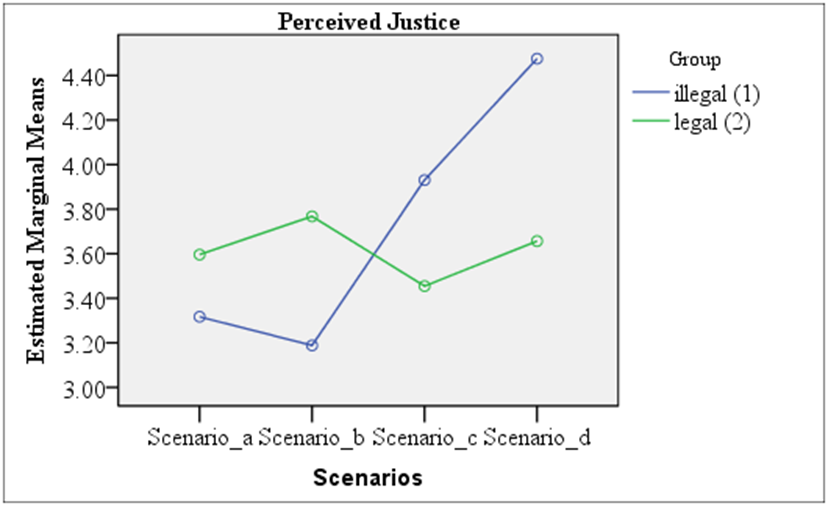}
  \caption{Justice}
  \label{fig:justice}
  \Description{Interaction effect between scenarios and groups regarding perceived justice. The starting point on the top is the legal violation group.}
\end{subfigure}
\begin{subfigure}{0.33\textwidth}

  \includegraphics[width= \linewidth]{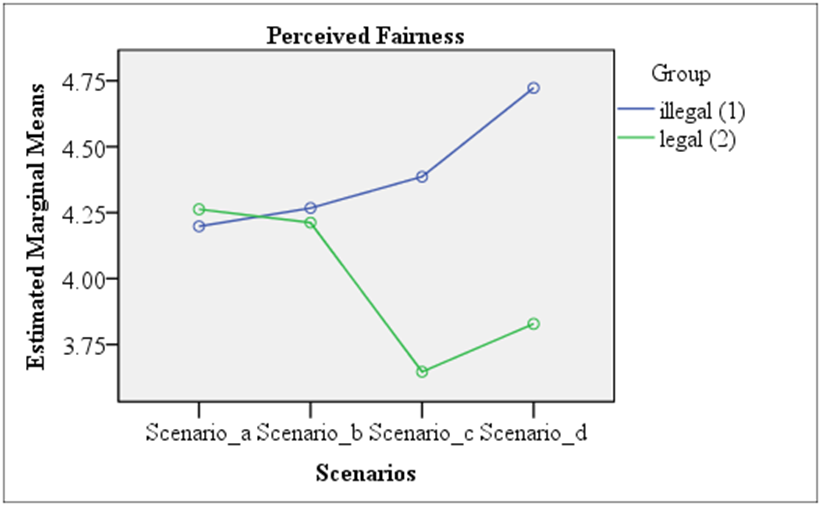}
  \caption{Fairness}
  \label{fig:fairness}
  \Description{Interaction effect between scenarios and groups regarding perceived fairness. the starting point on the top is the legal violation group.}
\end{subfigure}
 
\caption {Interaction between scenarios and groups regarding perceived justice and fairness.}
\end{figure*}

\subsection{Experimental Design}

Our online experiment used a between-subjects design by legality (Group 1 illegal, Group 2 legal). Then, we used a within-subjects design for each group with four scenarios (1a, 1b, 1c, 1d, 2a, 2b, 2c, 2d). We showed each group a total of 4 mock scenarios designed by the team, as shown in \autoref{a:design}. The first group (Group 1) of participants viewed four scenarios in which illegal content was posted to Reddit and Twitter. The second group viewed four scenarios where legal content was posted to Reddit and Twitter, as shown in \autoref{tab:scenario}. We ran a series of mixed ANOVA with Bonferroni post-hoc tests to answer RQ1 and test our main hypotheses, H2s and H3s. 

Before creating the scenarios, the team read Twitter’s Community Guidelines and Reddit’s Content Policy. We did this to create illegal and legal content from scratch that would violate the two platforms’ policies and would likely be removed from the platform. Team members wrote the illegal and legal content of all eight scenarios for the online experiment design. We presented the scenarios in the lab meeting to collect feedback and modify them. The illegal scenarios are consistently about terrorism inducing public safety concerns, and legal scenarios are consistently about sexism and racism targeting a specific group or individual. 
Each scenario states either a Reddit or Twitter user posted the message or tweet to either Reddit or Twitter. The scenario then states the message or tweet the fake Reddit or Twitter user posted. Following this, the scenario states the removal from the platform by the moderator on the respective platforms for violating the platform’s policy. The scenario states the policy that the message or tweet violated. Additionally, the scenario states the moderator either warned the user or permanently banned the user. After each scenario, participants answered questions about perceived justice and fairness. Scenarios and survey questions are in \autoref{a:scenario}.

\subsection{Measurements}

The participants’ perception of justice for each scenario was measured with a single item: “Is it necessary to punish the (Reddit or Twitter) user for their post to deliver justice?”. Participants answered on a 5-point Likert scale from 1 (Absolutely Not Necessary) to 5 (Absolutely Necessary). This item was extracted and adapted from the Punishment Orientation Questionnaire \cite{yamamoto_creating_2019}. 
The participants’ perception of fairness for each scenario was measured with a single item: “How fair do you perceive the (Reddit or Twitter) moderator’s decision to be?”. Participants answered on a 5-point Likert scale from 1 (very unfair) to 5 (very fair).

\section{Results}

\subsection{Perceived Justice}

Mauchly’s Test of Sphericity indicated that sphericity was met (\textit{W} = .98,  $\chi^2(5)$ = 4.02, \textit{p} = .547). The tests of within-subjects effects show a significant main effect among the scenarios (\textit{F}(3,594) = 21.45, \textit{p} < .001, $\eta^2$ = .10). There was also a significant interaction between scenarios and groups (\textit{F}(3,594) = 21.05, \textit{p} < .001, $\eta^2$ = .13). The partial eta squared indicated that interaction had a stronger impact than the main effect on justice. All the descriptive statistics for each scenario and the \textit{p} values between groups are shown in \autoref{tab:justice}. The interaction effect is shown in  \autoref{fig:justice}.

RQ1 tries to understand if users’ perceived justice was higher or lower for user-moderated platforms (i.e., Reddit) versus commercially moderated platforms (i.e., Twitter); we need to specifically compare scenario\_1a and 1b, scenario\_1c and 1d, scenario\_2a and 2b, and scenario\_2c and 2d. The Bonferroni posthoc tests with simple effects showed a perceived justice difference in scenario\_1c and 1d ( \textit{p} < .001). Overall, perceived justice is higher for commercially moderated than for user-moderated platforms in illegal scenarios with retributive moderation strategies. 

H2a stated that perceived justice is higher in illegal than legal scenarios. The tests of between-subjects effects showed that, in general, there is no significant difference in perceived justice between illegal and legal scenarios (\textit{p} = .487). Bonferroni posthoc tests comparing justice in scenarios revealed a significant difference in scenarios\_b, c, and d; specifically, in scenario\_b, perceived justice was lower in the illegal group compared to the legal group (\textit{p} = .002); in scenario\_c,  perceived justice is higher in the illegal group compared to the legal group (\textit{p} = .019); and in scenario\_d, perceived justice is higher in the illegal group compared to the legal group  (\textit{p} < .001). Overall, the perceived justice was higher in illegal scenarios compared to legal scenarios with retributive moderation strategies and had either no difference or was lower in illegal scenarios compared to legal scenarios with restorative moderation strategies. Thus, H2a was partially supported. 

H3a stated that perceived justice was higher for retributive moderation strategies than restorative moderation strategies. Specifically, we compare scenario\_a and c to see the different moderation strategies on commercially moderated platforms and compare scenario\_b and d to see the different moderation strategies on user-moderated platforms. The simple effects for the interaction revealed that the perceived justice increased significantly from scenario\_a to scenario\_c in the illegal group (\textit{p} < .001) but showed no significant difference in the legal group (\textit{p} = 1.00). The simple effects for the interaction revealed that the perceived justice increased significantly from scenario\_b to scenario\_d in the illegal group (\textit{p} < .001) but showed no significant difference in the legal group (\textit{p} = 1.00). 
Overall, the perceived justice was higher for retributive compared to restorative moderation strategies in the illegal group but had no difference in the legal group.  H3a was partially supported. In other words, in illegal scenarios, retributive moderation is more just; in legal scenarios, there is no justice difference regarding retributive/restorative moderation. 

\subsection{Perceived Fairness}

Mauchly’s Test of Sphericity indicated that sphericity was not met (\textit{W} = .79, $\chi^2(5)$ = 46.50, \textit{p} < .001). The tests of within-subjects effects show a significant main effect among the scenarios (\textit{F}(2.59, 512.25) = 4.16, \textit{p} = .009, $\eta^2$ = .02). There was also a significant interaction between scenarios and groups (\textit{F}(2.59, 512.25) = 17.38, \textit{p} < .001, $\eta^2$ = .08). The partial eta squared indicated that interaction had a stronger impact than scenarios on fairness. All the descriptive statistics for each scenario and the \textit{p} values between groups are shown in \autoref{tab:fairness}. The interaction effect is shown in \autoref{fig:fairness}.

 \begin{table*}[h]
 \small\sf\centering
\caption{Descriptive Statistics for Each Group and Scenario in Terms of Perceived Fairness}
\label{tab:fairness}
\begin{tabular}{llllll}
\toprule
                & Fairness-a          & Fairness-b          & Fairness-c          & Fairness-d          & Fairness-overall    \\ \midrule
                & Scenario\_1a        & Scenario\_1b        & Scenario\_1c        & Scenario\_1d        &                     \\
Group1: Illegal & \textit{M} = 4.20, \textit{SD} = 1.00 & \textit{M} = 4.27, \textit{SD} = 0.96 & \textit{M} = 4.39, \textit{SD} = 0.99 & \textit{M} = 4.72, \textit{SD} = 0.74 & \textit{M} = 4.39, \textit{SE} = 0.08 \\ \midrule
                & Scenario\_2a        & Scenario\_2b        & Scenario\_2c        & Scenario\_2d        &                     \\
Group2: Legal   & \textit{M} = 4.26, \textit{SD} = 0.89 & \textit{M} = 4.21, \textit{SD} = 1.01 & \textit{M} = 3.65, \textit{SD} = 1.39 & \textit{M} = 3.83, \textit{SD} = 1.33 & \textit{M} = 3.99, \textit{SE} = 0.08 \\                       \midrule
                & \textit{p} = .630           & \textit{p} = .692           & \textit{p} <  .001   & \textit{p} <  .001   & \textit{p} <  .001   \\ \bottomrule
\end{tabular}%
\end{table*}


RQ1 tries to understand if users’ perceived fairness was higher or lower for user-moderated platforms versus commercially moderated platforms. We specifically compare scenario\_1a and 1b, scenario\_1c and 1d, scenario\_2a and 2b, and scenario\_2c and 2d.  The Bonferroni posthoc tests with simple effects showed only a perceived fairness difference in scenario\_1c and 1d (\textit{p} < .001). Overall, perceived fairness is higher for commercially moderated platforms than for user-moderated platforms in illegal scenarios with retributive moderation strategies. 

H2b stated that perceived fairness was higher in illegal than legal scenarios. The tests of between-subjects effects showed that, in general, there was a significant difference in perceived fairness between illegal and legal scenarios (\textit{p} < .001).  Bonferroni posthoc tests comparing fairness in scenarios revealed a significant difference in scenarios\_c and d; specifically, in scenario\_c, perceived fairness is higher in the illegal group compared to the legal group (\textit{p} < .001), and in scenario\_d, perceived fairness is higher in the illegal group compared to the legal group (\textit{p} < .001). Overall, the perceived fairness was higher in illegal scenarios compared to legal scenarios with retributive moderation strategies, and there was no difference in illegal scenarios compared to legal scenarios with restorative moderation strategies. Thus, H2b was partially supported. In other words, moderating illegal scenarios was fairer than legal scenarios in scenarios with retributive moderation, but there was no fairness difference regarding legality in scenarios with restorative moderation.  

H3b stated that perceived fairness is higher for retributive than restorative moderation strategies. We compared scenario\_a and c to see the different moderation strategies on commercially moderated platforms and compared scenario\_b and d to see the different moderation strategies on user-moderated platforms. The simple effects revealed that perceived fairness showed no significant difference in the illegal group (\textit{p} = .760) but decreased significantly in the legal group (\textit{p} < .001) from scenario\_a to c. The simple effects revealed that the perceived fairness increased significantly from scenario\_b to d in the illegal group (\textit{p} < .001) but decreased significantly in the legal group (\textit{p} = .012). Overall, the perceived fairness was higher or no difference for retributive compared to restorative moderation strategies in the illegal group but lower in the legal group. H3b was partially supported. In other words, retributive moderation is fairer in illegal scenarios, but in legal scenarios, restorative moderation is fairer.  

\section{Discussion}

\subsection{Not Only Restorative But Retributive Moderation Can also Improve Justice and Fairness}
Recent research proposes to move from retributive justice to restorative justice to mediate the harm and resolve the conflict between violators and victims \cite{,schoenebeck_reimagining_2021}. This line of research criticizes that retributive justice can even cause severe sequential harm to the stakeholders \cite{schoenebeck_drawing_2021,schoenebeck_youth_2021}. However, they often consider online toxicity as a whole and do not consider the nuanced difference of violation types. In this study, we separate violation from the legality perspective. Our results consistently show that users consider retributive moderation strategies to deliver higher justice and fairness for illegal violations than legal ones, and restorative strategies deliver higher justice for legal violations than illegal ones.   In this sense, we contribute to a nuanced understanding of violation and perceived justice differences. First, we supplement the line of restorative justice research \cite{schoenebeck_drawing_2021} and point out that it is more about the legal violation related to online toxicity and harassment. 

Second, retributive moderation is still preferred to deal with illegal violations. Users still think it is fair for illegal violations to receive severe punishment, such as terrorism and public violence. Speculatively, an illegal violation is more likely to cause severe consequences for public safety or society, like content inflaming civil unrest or terrorism. These types of violations should be punished severely online or even have the police involved offline. Legal violation is more likely to cause psychological harm for individual entities, like harassment towards a specific group related to gender, race, identity, or disability \cite{lyu_i_2024,uttarapong_harassment_2021}.
Consequently, we suggest that both types of moderation strategies should be incorporated into the platform’s moderation policymaking, and it is difficult to weigh which one is better and should be emphasized. Policymakers should try to differentiate the types of violations. Though the potential legal violation can sometimes lead to illegal violations at scale, it is essential to explore the transition between these two types of violations further and identify the appropriate boundary to intervene with relevant agencies.

We can also get some clues from the tech giants’ moderation policies about their different attitudes toward legal or illegal violations. For example, Facebook, Microsoft, Twitter, and YouTube established the “Global Internet Forum to Counter Terrorism” in 2017 to coordinate content removal about “violent terrorist imagery and propaganda”. However, there is little or no collaboration about daily online harassment. Different platforms have different policies regarding online harassment, such as community guidelines and codes of conduct. Such situations also cause challenges from the policymaking at the platform level and hinder platform collaboration. It seems there are no clear collaborative solutions for the legal violations since different platforms hold different values, such as the Stormfront white nationalist website \cite{hemphill_very_2022}. This study also sheds light on collaborative policymaking to deal with legal violations from some commonly shared values for most platforms, such as racism and sexism, with zero tolerance. 

\subsection{Explanation Can Improve Justice and Fairness in Certain Scenarios But Not All}

In line with work about moderation explanation increasing perceived justice and fairness \cite{jhaver_did_2019}, we extend and show how the explanation of moderation decisions across different platforms affects users’ perceived justice and fairness. Regarding illegal violations, retributive strategies on commercially moderated platforms deliver higher justice and fairness than user-moderated platforms. Many scholarships have highlighted the importance of moderation transparency to improve justice and fairness \cite{felzmann_towards_2020, suzor_what_2019}. Most of them focus on one platform’s moderation policies. For example, many criticize that commercial moderation lacks transparency and that community moderation should keep increasing transparency with clear rules and norms with active engagement \cite{ma_transparency_2023,ma_im_2022}. We contribute to the comparison of these types of moderation with an explanation. Users’ perception of the platform difference after viewing explanations also plays a significant role in perceived justice and fairness only in illegal violations with retributive moderation. Offering explanations with restorative moderation strategies seems to have no platform difference, even retributive moderation under legal scenarios. In this sense, we provide a nuanced understanding of the moderation strategies in different scenarios with platform differences and show that transparency is essential to justice and fairness in a specific situation, but not all. Such findings supplement prior work regarding the tension of punishing and nurturing \cite{jiang_trade-off-centered_2023} and offer naunaced differences to consider moderation resource allocation. For example, with scenarios where the explanation makes no or little difference, the platform should invest less human labor and resources.

\section{Limitation and Future Work}
Although this study has important implications for content moderation policies, it is subject to a few limitations. Firstly, since many people in the US come from diverse cultures, their cultural biases may be influenced by living or growing up in the US. Thus, people from the same culture in the US may hold different cultural views compared to those in their country of origin. Therefore, future studies should take into account the cultural differences in users of different countries \cite{jiang_understanding_2021} and investigate how different perceptions of social media use in other regions affect content moderation. Secondly, we used a single-item measurement for fairness and justice, whereas most other fairness and justice research employs qualitative methods. It is crucial for HCI scholars to develop a scale for future content moderation studies. Additionally, the preliminary study did not explore action variables, so it would be beneficial to investigate how users would like platforms to take action regarding content moderation to make their decisions more just and fair. Last, there are also potential opportunities  to explore how to incorporate the policy into moderation system design \cite{yang_designing_2023} and to reshape the governance and platform identity to make the platform inclusive and diverse
\cite{das_jol_2021}.  

\begin{acks}
This research was funded by the National Science Foundation (Award No. 1928627).
\end{acks}

\balance

\bibliographystyle{ACM-Reference-Format}
\bibliography{references}


\appendix

\section{Experimental Design Process}
\label{a:design}
 \begin{figure*}[htbp]
  \includegraphics[width=\textwidth]{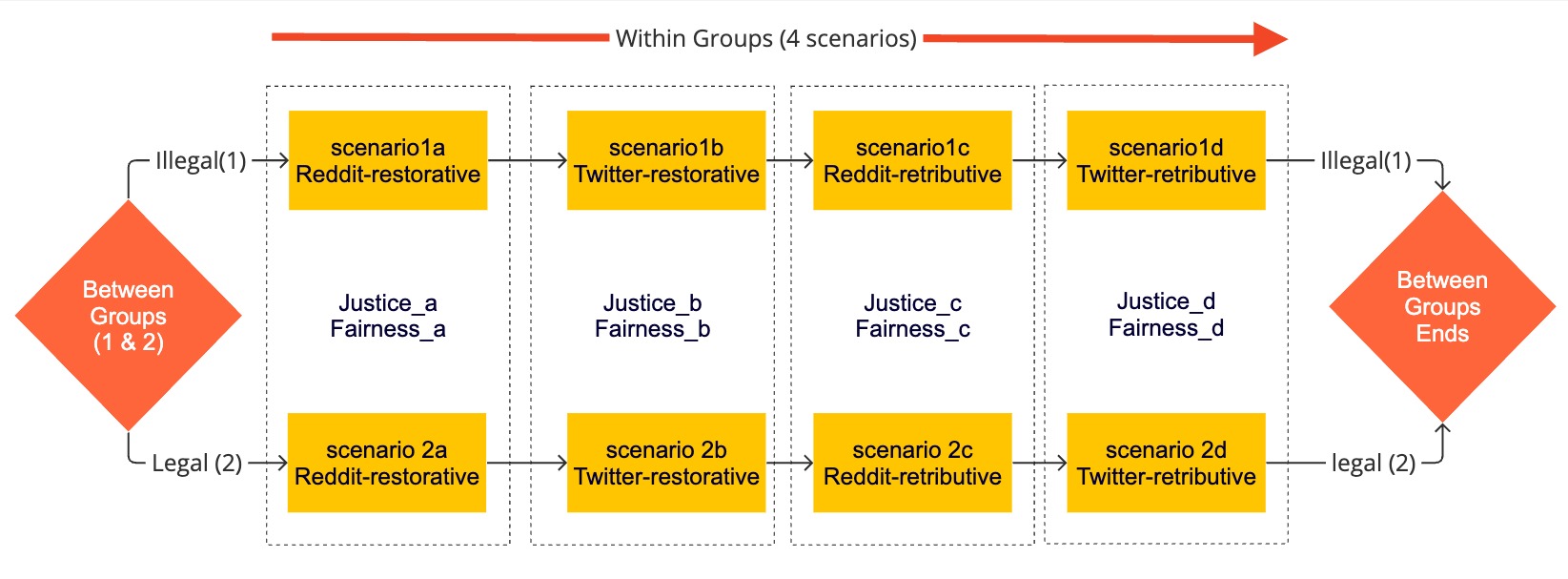}
  \caption{Experimental design process.}
  \label{fig:process}
  \Description{This is a figure to show the experimental design process with two groups of between-subjects design first, and then within each group, there is a repeated measure with four different scenarios.}
\end{figure*}

 \newpage

\section{Scenarios and Main Questions}
\label{a:scenario}

\begin{table}[h]
\label{tab:my-table}
\begin{tabular}{
>{\columncolor[HTML]{FFFFFF}}c }
\toprule
{\color[HTML]{FE0000} \textbf{Warning: The following content might be disturbing.}} \\ \bottomrule
\end{tabular}
\end{table}

\subsection{Illegal Scenarios}

\subsubsection{Scenario 1a – Restorative, User, Illegal}
A Reddit user posted this message in a subreddit.

\begin{quote}
    \textit{Dear god these fucking snowflakes protest everything. They’re such idiots. About to start pulling up to their marches and blow them up to teach them a lesson. Who’s coming with?}
\end{quote}

This user’s message was removed by the moderators of the subreddit. This post was removed for 
violating Reddit’s Content Policy Rule 7: “Keep it legal, and avoid posting illegal content…”. 
Instigating terrorism is illegal. In addition, the moderators gave the user\textbf{ a warning} saying if they 
post any more illegal content, \textbf{they will be banned from the subreddit}.

How fair do you perceive the Reddit moderator's decision to be? (The question will repeat after each scenario)
\begin{itemize}
\item Very Unfair
\item Unfair
\item Neutral
\item Fair
\item Very Fair
\end{itemize}


Is it necessary to punish the Reddit user for their post to deliver justice? (The question will repeat after each scenario)
\begin{itemize}
\item Absolutely Not Necessary
\item Slightly Necessary
\item Moderately Necessary
\item Very Necessary
\item Absolutely Necessary
\end{itemize}

\subsubsection{Scenario 1b – Restorative, Commercial, Illegal}
A Twitter user posted this tweet to their account.

\begin{quote}
    \textit{Got a new supply of Havendust. Shoot me a DM if you wanna try some of this new icing. Local pickup only.}
\end{quote}

This user’s tweet was removed from the platform by Twitter’s moderators. This tweet was 
removed for violating Twitter’s Community Guidelines that state “You may not use our service 
for any unlawful purpose or in furtherance of illegal activities. This includes selling, buying, or 
facilitating transactions in illegal goods or services, as well as certain types of regulated goods or 
services.” Selling controlled substances is illegal. In addition, the user was also given \textbf{a warning }
saying if they post any more illegal content, their \textbf{account will be suspended} from posting on 
Twitter.

\subsubsection{Scenario 1c – Retributive, User, Illegal}
A Reddit user posted this message in a subreddit.

\begin{quote}
    \textit{If one more fucking leftie posts a damn petition about some stupid ass leftie problem, my gun is coming outta the safe and a bullet is going through their head. Gonna call my hunting buddies to come with me too. Come join us in front of City Hall at 1pm.}
\end{quote}

This user’s message was removed by the moderators of the subreddit. This post was removed for 
violating Reddit’s Content Policy Rule 7: “Keep it legal, and avoid posting illegal content…”. 
Instigating terrorism is illegal. In addition, the moderators \textbf{permanently banned} the user from 
posting in the subreddit.

\subsubsection{Scenario 1d – Retributive, Commercial, Illegal}
A Twitter user posted this tweet to their account.

\begin{quote}
    \textit{Took sweet ass bath pictures of a 2-year-old when I was babysitting. DM me if you want to see them. Will sell to the highest bidder.}
\end{quote}

This user’s tweet was removed from the platform by Twitter’s moderators. This tweet was 
removed for violating Twitter’s Community Guidelines that state “We have zero tolerance for 
child sexual exploitation on Twitter.” Child sexual exploitation is illegal. In addition, the 
moderator \textbf{permanently banned} the user’s account from posting on Twitter.

\subsection{Legal Scenarios}

\subsubsection{Scenario 2a – Restorative, User, Legal}
A Reddit user posted this message in a subreddit.
\begin{quote}
    \textit{Gay people are heathens. They don’t deserve to live and spread their disgustingness. Proud of the guy who shot up that gay nightclub. They deserved to die.}
\end{quote}

This user’s message was removed by the moderators of the subreddit. This post was removed for 
violating Reddit’s Content Policy Rule 1: “Remember the human. Reddit is a place for creating 
community and belonging, not for attacking marginalized or vulnerable groups of people. 
Everyone has a right to use Reddit free of harassment, bullying, and threats of violence. 
Communities and users that incite violence or that promote hate based on identity or vulnerability 
will be banned.” In addition, the moderators gave the user \textbf{a warning} saying if they post any more 
hateful content, \textbf{they will be banned from the subreddit}.

\subsubsection{Scenario 2b – Restorative, Commercial, Legal}
A Twitter user posted this tweet to their account.
\begin{quote}
    \textit{Feminists are disgusting. They are spreading their filthy ideas that women are just as good 
as men and must be stopped. I wouldn’t mind if they were all raped.}
\end{quote}

This user’s tweet was removed from the platform by Twitter’s moderators. This tweet was 
removed for violating Twitter’s Community Guidelines that state “You may not promote violence 
against, threaten, or harass other people on the basis of race, ethnicity, national origin, caste, sexual 
orientation, gender, gender identity, religious affiliation, age, disability, or serious disease.” In 
addition, the user was also given \textbf{a warning} saying if they post any more hateful content, their 
\textbf{account will be suspended} from posting on Twitter.

\subsubsection{Scenario 2c – Retributive, User, Legal}
A Reddit user posted this message in a subreddit.

\begin{quote}
    \textit{Damn leftie sheep believing in fake science and going around wearing masks. COVID isn’t real. Take off your masks or I’m gonna cough on you.}
\end{quote}

This user’s message was removed by the moderators of the subreddit. This post was removed for 
violating Reddit’s Content Policy Rule 1: “Remember the human. Reddit is a place for creating 
community and belonging, not for attacking marginalized or vulnerable groups of people. 
Everyone has a right to use Reddit free of harassment, bullying, and threats of violence. 
Communities and users that incite violence or that promote hate based on identity or vulnerability 
will be banned.” In addition, the moderators \textbf{permanently banned} the user from posting in the 
subreddit.

\subsubsection{Scenario 2d – Retributive, Commercial, Legal}
A Twitter user posted this tweet to their account.

\begin{quote}
    \textit{Mexican kids at the border should be tear-gassed. They’re coming into our country to take our jobs.}
\end{quote}

This user’s tweet was removed from the platform by Twitter’s moderators. This tweet was 
removed for violating Twitter’s Community Guidelines that state “You may not promote violence 
against, threaten, or harass other people on the basis of race, ethnicity, national origin, caste, sexual 
orientation, gender, gender identity, religious affiliation, age, disability, or serious disease.” In 
addition, the moderator \textbf{permanently banned} the user’s account from posting on Twitter.

\end{document}